\documentclass[aps,prl,twocolumn,a4paper,10pt]{revtex4-1}

\usepackage[T1]{fontenc}		
\usepackage[english]{babel}		
\usepackage[latin1]{inputenc}	
\usepackage{times}

\usepackage{amsmath}			
\usepackage{amssymb}			
\usepackage{bbm}				

\usepackage[colorlinks=true, a4paper=true, pdfstartview=FitV,linkcolor=blue, citecolor=blue, urlcolor=blue]{hyperref}

\usepackage{graphicx}			
\usepackage{graphics}			
\DeclareGraphicsExtensions{.pdf}
\usepackage{color}		

\newcommand{\bea}{\begin{eqnarray}}
\newcommand{\eea}{\end{eqnarray}}

\newcommand{\bk}{\mathbf{k}}

\newcommand{\bd}{\mathbf{d}}

\newcommand{\bsigma}{\boldsymbol{\sigma}}

\begin{document}

\title{Exceptional links and twisted Fermi ribbons in non-Hermitian systems}

\author{Johan~Carlstr\"om and Emil~J. Bergholtz}
\affiliation{Department of Physics, Stockholm University, 106 91 Stockholm, Sweden}
\date{\today}

\begin{abstract}
The generic nature of band touching points in three-dimensional band structures is at the heart of the rich phenomenology, topological stability, and novel Fermi arc surface states associated with Weyl semimetals. Here we report on the corresponding scenario emerging in systems effectively described by non-Hermitian Hamiltonians. Remarkably, three-dimensional non-Hermitian systems have generic band touchings along one-dimensional closed contours, forming exceptional rings and links in reciprocal space. The associated Seifert surfaces support open "Fermi ribbons" where the real part of the energy gap vanishes, providing a novel class of higher-dimensional bulk generalizations of Fermi arcs which are characterized by an integer twist number. These results have possible applications to a plethora of physical settings, ranging from mechanical systems and optical metamaterials with loss and gain to heavy fermion materials with finite-lifetime quasiparticles. In particular, photonic crystals provide fertile ground for simulating the exuberant phenomenology of exceptional links and their concomitant Fermi ribbons.

\end{abstract}
\maketitle

{\it Introduction.---}
A crucial insight that guides contemporary physics, is that central properties in nature can be understood in terms of topological arguments, that are in principle independent of the underlying details of the system in question \cite{PhysRevLett.49.405}. 
Recently, this point of view has experienced a dramatic surge, with applications to new forms of topological matter that admit emergent Weyl fermions accompanied by Fermi arcs  \cite{doi:10.1146/annurev-conmatphys-031016-025225,PhysRevB.83.205101,RevModPhys.90.015001,volovik2009universe}, as well as bulk insulating topological phases that are associated with metallic surface states \cite{RevModPhys.82.3045,RevModPhys.83.1057}. 
Yet the topological viewpoint is by no means new to physics, and emerges very naturally from the notion of a wave function. 
Likewise, in classical systems it is encountered via critical behavior and the proliferation of topological defects \cite{0022-3719-6-7-010}. 

In fact, topological arguments can be traced back to before the advent of modern physics, as exemplified by Lord Kelvin's proposal that atoms are in fact knots in the ether \cite{Kelvin}. Despite the subsequent rejection of this idea, as well as the entire notion of ether, the concept of emergent topological particlelike excitations has continued to motivate a search for field theories that admit stable knot solutions \cite{Skyrme127,Ranada1989,FaddevNiemi,PhysRevB.65.100512,Sutcliffe3001}. 

With the current focus on topological band structures, proposals have also emerged for the realization of knotlike structures in reciprocal space. Examples of this include Hopf insulators \cite{PhysRevLett.101.186805} as well as nodal link and knot semimetals \cite{2016Natur.538...75B,PhysRevB.96.041102,PhysRevB.96.041103,PhysRevB.96.081114,PhysRevB.96.201305} and superconductors \cite{PhysRevLett.119.147001},
though the gapless examples do face a fundamental challenge, as nodal lines are not generic in 3D systems \cite{PhysRev.52.365}. The most explored route around this problem is to rely on additional symmetries that protect the nodal lines. In particular, recent experiments on mirror-symmetric photonic crystals report observations of nodal chains in the spectrum \cite{NodalChain}. 

A second and less explored path to the realization of line nodes is to add dissipation to the system, which gives rise to an effective theory with non-Hermitian terms in the Hamiltonian \cite{Bender,2018arXiv180508200M}. Remarkably, this reduces the number of equations that describe band touching points from $3$ to $2$ \cite{Berry2004}, so that nodes become generically pointlike in 2D, and linelike in 3D. 
It should be noted though that unlike in Hermitian systems, these nodes are generally exceptional points where the Hamiltonian becomes defective and thus lacks a full spectrum of eigenvectors \cite{Berry2004,2018arXiv180508200M}. Furthermore, these exceptional points are connected by bulk Fermi arcs in 2D \cite{2017arXiv170805841K}, which were recently observed in experiments on photonics crystal slabs \cite{Zhoueaap9859}. Beyond photonics systems, experimental routes that have been proposed to realize a non-Hermitian band theory include dissipative ultracold atomic systems \cite{PhysRevLett.118.045701}, disordered Weyl semimetals \cite{PhysRevB.97.041203}, and heavy fermion systems that support finite-lifetime quasiparticles \cite{2018arXiv180501172Y}. 
While the field of non-Hermitian topological systems has seen an explosive development recently \cite{2018arXiv180508200M,2018arXiv180506492K,PhysRevB.97.045106,PhysRevLett.118.040401,PhysRevA.89.062102,PhysRevLett.120.146402,PhysRevA.93.062101,2015PhLA..379.1213Y,PhysRevLett.102.065703,PhysRevLett.120.146402,2018arXiv180409975W,PhysRevLett.115.200402,2018arXiv180301876Y,2399-6528-2-3-035043,PhysRevLett.116.133903}, the focus has been almost exclusively on one- and two-dimensional phenomena.

\begin{figure*}[!htb]
 \hbox to \linewidth{ \hss
\includegraphics[width=\linewidth]{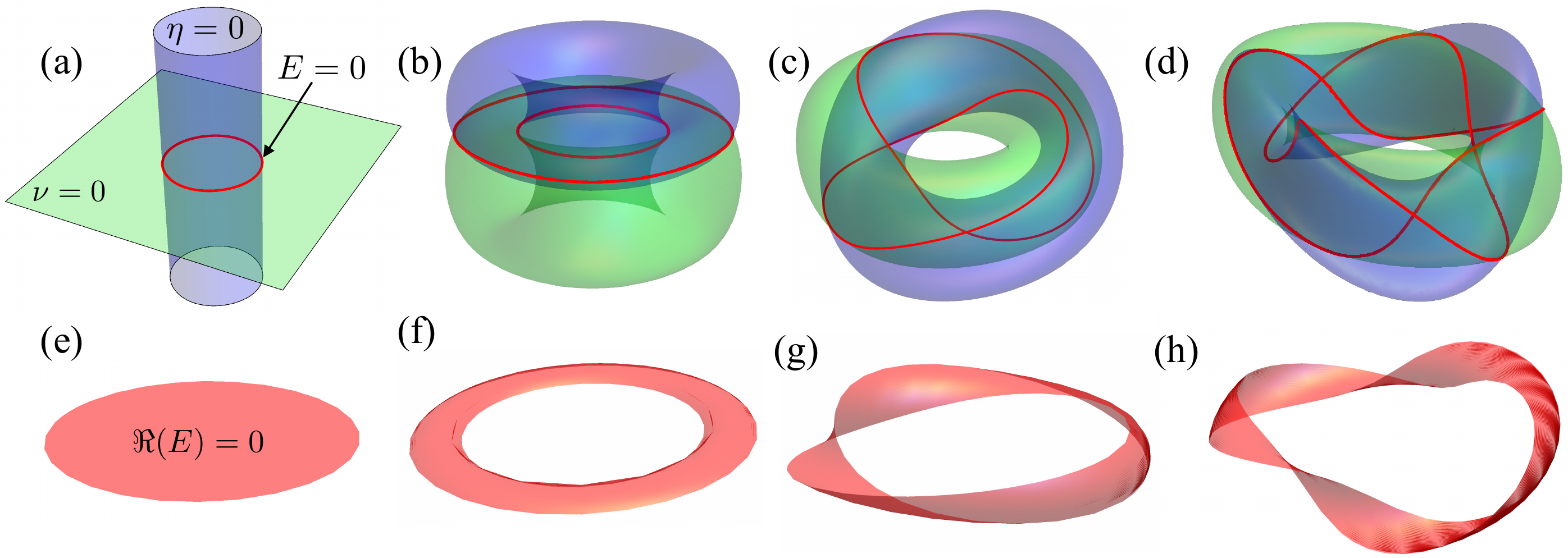}
 \hss}
\caption{
Exceptional line topology in non-Hermitian systems. The top row (a)--(d) shows the surfaces $\nu=0$ in green and $\eta=0$ in blue, see Eq. (\ref{nu}) for a definition. Their intersections give the exceptional lines where the bands meet and the theory becomes defective, here displayed in red.
When particle-hole symmetry is present, these correspond to $E=0$, i.e., zero energy and infinite lifetime. 
 The bottom row (e)-(h) shows the corresponding Fermi surface topology.
In (a) a non-Hermitian term is added to a single Weyl node so that it splits into an exceptional ring that is accompanied by a disk-shaped Fermi surface (e). 
In (b) $\eta,\nu$ are tori with a relative translation, resulting in two exceptional rings and a Fermi surface in the form of a punctured disk. 
If both tilting and translation is applied to one of the tori, then the exceptional rings form a Hopf link (c). Finally, elliptic deformation --- stretching in one direction and contracting in the other --- in combination with bending gives rise to a twice-braided link (d). 
The Fermi surfaces corresponding to intersecting tori
 are Seifert surfaces \cite{seifert} with integer twists (g)-(h). 
Thus, the Fermi surface connects the two exceptional lines, or equivalently, the exceptional lines are terminations points of the now-open Fermi surfaces, which we dub "Fermi ribbons."
The exact form of the models (b)-(d) is outlined
in Eqs. (\ref{d0TB})--(\ref{bd3}).
}
\label{nodes}
\end{figure*}

In this paper we explore new striking phenomena that are unique to three-dimensional non-Hermitian systems. In particular, we predict the emergence of generic exceptional lines with a nontrivial topology encoded by a finite linking number [Figs. \ref{nodes}(a) -- \ref{nodes} (d)]. These are accompanied by generalizations of the aforementioned bulk Fermi arcs in the form of "Fermi ribbons"---open orientable surfaces of nontrivial topology as displayed in Figs. \ref{nodes} (e) -- \ref{nodes} (h). 
We also show that these systems can be realized by the addition of a non-Hermitian term to relatively simple tight-binding models and that these phenomena are within the reach of current experimental approaches.

{\it Defective Hamiltonians and open Fermi surfaces.---}
Assuming an interaction-free two-band system with particle-hole symmetry, we can write the Hamiltonian of the form
\bea 
H=\bd(\bk)\cdot \bsigma, \;\bd\in \mathbb{C}^3. \label{H}
\eea
Decomposing the $\bd$ vector in its real and imaginary parts according to $\bd=\bd_R+i \bd_I $, we find eigenvalues and thus energies of the form
\bea
E^2=\bd_R^2-\bd_I^2+2i \bd_R\cdot \bd_I .\label{eV2}
\eea
A striking implication of Eq. (\ref{eV2}) is that $\bd_R$ does not need to be zero anywhere in order for the system to possess band touching points ($E=0$). These can in principle be created by adding a non-Hermitian term to an insulator. 

Next we note that the nodal points are described by solutions to the equations
\bea\nonumber
 \bd_R^2-\bd_I^2=\eta=0,\\
\bd_R\cdot \bd_I=\nu=0\label{nu}. 
\eea
Since the system is particle-hole symmetric the band touching points necessarily correspond to zero energy modes with infinite lifetime. Furthermore, if $\bd\not=0$, then Eq. (\ref{nu}) does not describe conventional zeros of the Hamiltonian, but rather exceptional points where the theory becomes defective, i.e., where the Hamiltonian in Eq. (\ref{H}) lacks a complete basis of eigenvectors \cite{2018arXiv180508200M}. 

In two-dimensional systems, Eq. (\ref{nu}) describes closed lines in $\bk$ space where $\eta,\nu$ change sign. The intersections of these are thus generally pointlike and appear in pairs that are connected by bulk Fermi arcs \cite{2017arXiv170805841K}. In contrast to nodal points in Weyl semimetals and graphene however, they do not feature a linear spectrum but rather a dispersion of the form $E\sim \pm\sqrt{|\bk|}$ and thus divergent Fermi velocity \cite{2018arXiv180508200M}.

In three dimensions, Eq. (\ref{nu}) instead defines closed surfaces, with intersections that necessarily are closed lines in $\bk$ space \cite{PhysRevLett.118.045701}. 
In order for the eigenvalues of the spectrum to lie on the Fermi surface, i.e., to have a real part that is zero, 
Eq. (\ref{eV2}) implies
\bea
 \Re(E)=0\implies E\in \mathbb{I}\implies \nu=0,\;
 \eta\le 0 .\;\;\;\label{FermiSurface}
\eea
By contrast, if we require that the spectrum is real, we obtain the condition
\bea
 \Im(E)=0\implies E\in \mathbb{R}\implies \nu=0,\;
 \eta\ge 0 .\;\;\;\label{FermiSurface2}
\eea
Thus we see that the surface $\nu=0$ in principle can be decomposed into parts with either zero energy, or a purely real spectrum, according to
\bea\nonumber					
S_{1}: \;\;\;\nu=0,\;\eta\le 0,\;\Re(E)=0\\\nonumber
S_{2}: \;\;\;\nu=0,\;\eta\ge 0,\;\Im(E)=0 \\
S_{\nu=0}= S_1 \cup S_2 .\label{Si}
\eea
Furthermore, it follows from Eq. (\ref{eV2}) that $S_1$ is orientable, provided that $\eta<0$. Expanding $E$ in $\nu$ near $\nu=0$ we find 
\bea
E_\pm=\pm\sqrt{\eta+2i\nu}\approx \pm i\sqrt{|\eta|} \pm \frac{\nu}{\sqrt{|\eta|}},\label{orient}
\eea
so that $\Re(E_\pm)$ are to lowest order odd in $\nu$ as should be expected on the Fermi surface. 
However, in sharp contrast to Hermitian theories, defective models do necessarily possess a Fermi surface that is open, as can be seen from the following reasoning.
In the general case we have that $\nu=0$ on a set of closed surfaces: 
\bea
S_{\nu=0}=\sum_i S_{\nu=0}^i \label{Sdecomp}.
\eea
According to Eq. (\ref{nu}), the exceptional points form closed lines that lie on such a surface, so that $S_{\nu=0}^i$ is divided into subsets as described by Eq. (\ref{Si}). The implication of this is that the Fermi surface $S_1$ contains an element $S_1^i$ that is a part of a closed surface $S_{\nu=0}^i$, and which is correspondingly open. Thus, the exceptional lines should be interpreted as singularities where the Fermi surface terminates. These open Fermi surfaces provide an immediate generalization of the bulk Fermi arcs that were recently observed in non-Hermitian 2D systems \cite{Zhoueaap9859}.

{\it Topology of exceptional lines and Fermi ribbons.---}
By adding a dissipative term to a Weyl semimetal, we generally split Weyl nodes into exceptional rings \cite{PhysRevLett.118.045701,PhysRevB.97.075128}. 
From Eq. (\ref{eV2}) and the subsequent discussion, we thus conclude that there are two principal types of nodal rings: those that originate in Weyl nodes, and thus are topologically protected, and their trivial counterpart, which can in principle be gapped out by perturbations without merging with a partner of opposite topological charge.

According to the defectivity-Fermi surface relation outlined above, it follows that the splitting of a Weyl point into a nodal line is associated with the creation of an open Fermi surface that terminates on this line, see Figs. [\ref{nodes} (a) and (e)]. 

\begin{figure}[!htb]
 \hbox to \linewidth{ \hss
\includegraphics[width=\linewidth]{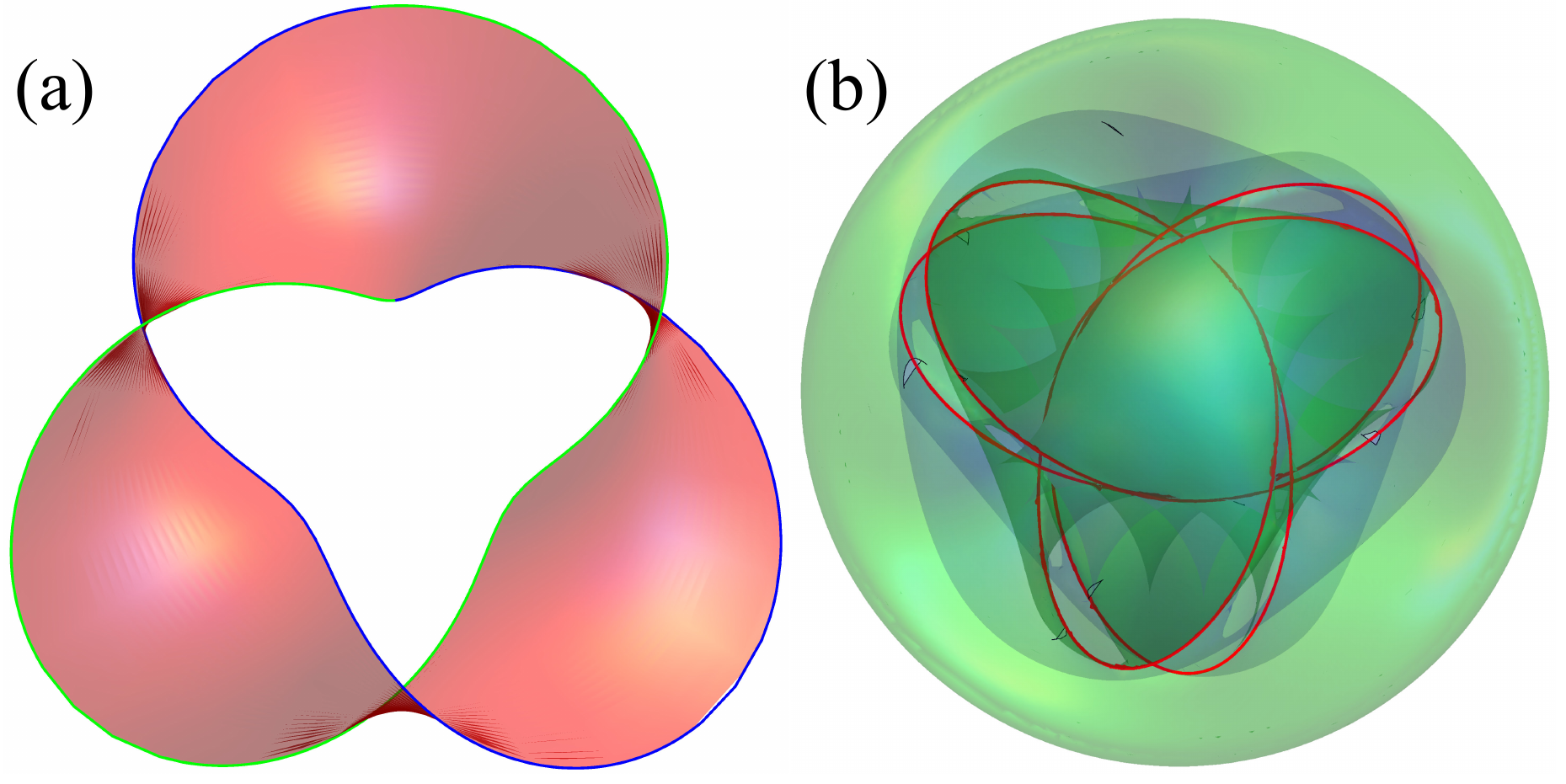}
 \hss}
\caption{
The three half-twist Möbius Fermi surface which terminates on a trefoil knot (a) is not an allowed nodal structure because it is not orientable. Still, these considerations do not rule out knots constructed from pairs of exceptional lines (b). In this case, it is a ribbon-shaped orientable Fermi surface, connecting the line-pair which forms a trefoil knot. This dispersion is described by Eqs. (\ref{knotEq}) and (\ref{dKnot}).
}
\label{threefoil}
\end{figure}

Furthermore, pursuing the preceding discussion to its logical conclusions, it is clear that the realization of topologically nontrivial exceptional objects faces a fundamental constraint in that the nodal lines must be compatible with a well-defined orientable Fermi surface that does not intersect itself. 
In particular, this rules out nodal structures consisting of a Möbius surface accompanied by an exceptional knot, as these are not orientable, as illustrated by Fig. \ref{threefoil} (a).
However, the preceding logics do not prevent pairs of exceptional lines forming a corresponding object, in which case it is an open Fermi surface in the form of a ribbon that forms the knot, as displayed in Fig. \ref{threefoil} (b).
Furthermore, it should be noted that exceptional knots may still in principle be constructed from substantially more complex Fermi surfaces than twisted ribbons.

In contrast to Möbius surfaces, Seifert surfaces \cite{seifert} that terminate on exceptional links are clearly compatible with the constraint of orientability, and should in principle be possible to realize in non-Hermitian tight-binding models.
Notably, Hopf links and higher-order generalizations thereof can essentially be constructed as intersections of tori, and so this provides a guiding principle for the geometry of $\eta,\nu$.

To construct concrete examples of models with nontrivial nodal topologies 
we consider a non-Hermitian term of the form  
\bea
\bd_I=\epsilon e_z, \label{dissipation}
\eea 
where $\epsilon$ is real and constant and $\{e_i\}$ denote basis vectors. This dissipative term is particularly relevant to experiments and has been recently realised in photonic parity-time-symmetric crystals \cite{10.1038/nmat4811}. However, it is clear from Eq. (\ref{eV2}) that the exceptional lines are invariant under simultaneous rotation of $\bd_R$, $\bd_I$, and so our argument has a high degree of generality. According to Eq. (\ref{nu}) the surface $\nu=0$ corresponds to $d_{R,z}=0$, suggesting that the first torus should be encoded directly in this vector component. Meanwhile, the second torus is to be imprinted into the $x-y$ components, giving a starting point of the form
\bea\nonumber
\bd_{R,0}= k_z e_x + (k_x^2+k_y^2-c)e_y\\
+ \big[-\epsilon^2+k_z^2+(k_x^2+k_y^2-c)^2 \big]e_z.\label{tori0}
\eea
Here, $d_{x,0}^2+d_{y,0}^2=\epsilon^2$ and $d_{z,0}=0$ describe a pair of tori while $\{e_i\}$ are basis vectors. The realization of nontrivial nodal topologies can then be achieved by perturbing $\mathbf{d}_0$, see Fig. \ref{nodes}, and subsequently encoding the resulting dispersion in a tight-binding model. 

Applying a translation in the $k_z$ direction to one of the tori, the intersection will take the form of two disconnected rings [Fig. \ref{nodes} (b)] accompanied by a Fermi surface shaped as a punctured disk [Fig. \ref{nodes} (f)]. 
Adding tilt to the torus, the exceptional lines becomes braided and form a Hopf link, see Fig. \ref{nodes} (c). The Fermi surface then becomes a twisted Seifert surface as illustrated in Fig. \ref{nodes} (g). Higher linking numbers can be achieved through more complex deformations of the torus: stretching and contracting it to an elliptic shape followed by bending results in exceptional lines that are twice braided  [see Fig. \ref{nodes} (d)] and a Fermi surface with two full twists [Fig. \ref{nodes} (h)].

{\it Realisation in tight-binding models.---} 
To construct theories with nontrivial nodal topologies in the context of a lattice model, we note that it is possible to encode a $\bd$ vector that is closely similar to Eq. (\ref{tori0}) as follows:
\bea\nonumber
\bd_{R,0}=e_x\sqrt{2}\sin k_z  +\Big(\cos k_x+\cos k_y-\frac{3}{2}\Big)e_y\;\;\;\\
+\Big[4-\epsilon^2-4\cos k_z +\Big(\cos k_x+\cos k_y-\frac{3}{2}\Big)^2  \Big]e_z.\;\;\;\label{d0TB}
\eea
A translation in the $k_z$ direction then corresponds to $k_z\to k_z+\Delta_z$. Expanding $\sin (k_z+\Delta_z)$ to first order in $\Delta_z$ we find a correction to the $\bd$ vector of the form
\bea
\delta \bd_1=e_x\sqrt{2} \;\Delta_z \;\cos k_z.\label{bd1}
\eea
Explicitly choosing $\Delta_z=-1/3$ and $\epsilon=\sqrt{1/8}$, we obtain the system shown in Fig. \ref{nodes} (b).
 
To construct a Hopf link, we need to first apply a tilt to Eq. (\ref{d0TB}), given by $\sqrt{2}\sin k_z\to \sqrt{2}(\sin k_z+t \sin k_y)$. Translation in the $k_x$ direction takes the form $\cos k_x\to\cos( k_x+\Delta_x)$. To the lowest order this gives 
 \bea
 \delta \bd_2=e_xt\; \sqrt{2}\sin k_y -e_y\Delta_x  \;\sin k_x .\label{bd2}
 \eea
 Taking $t=-\pi/16$, $\Delta_x=-1/4$, and $\epsilon=\sqrt{1/8}$ results in the dispersion shown in Fig. \ref{nodes} (c).
 
 To obtain exceptional lines that braid twice, we first bend one of the tori in Eq. (\ref{d0TB}) by including a term of the form $\sim [\cos(k_x-k_y)-\cos(k_x+k_y)]e_x$ and then introduce an elliptic deformation according to $[\cos k_x-\cos k_y]e_y$. This gives a correction of the form
 \bea\nonumber
 \delta \bd_3= b  [\cos(k_x-k_y)-\cos(k_x+k_y)]e_x\\
 +l[\cos k_x-\cos k_y]e_y,\label{bd3}
 \eea
where $b=-0.15$ and $l=0.3$ results in the dispersion shown in Fig. \ref{nodes} (d). 

Next, we note that the $e_z$ component of $\bd_{R,0}$ in Eq. (\ref{d0TB}) takes the explicit form 
\bea\nonumber
\Big(\cos k_x+\cos k_y-\frac{3}{2}\Big)^2
=\frac{9}{4}+\frac{1}{2}(2+\cos [2k_x]+\cos [2k_y])\\\nonumber
+\cos(k_x-k_y)+\cos(k_x+k_y)-3 (\cos k_x+\cos k_y)
\eea
and so the essential prerequisite for realizing the nodal structures illustrated in Fig. \ref{nodes} and defined in Eqs. (\ref{d0TB})-(\ref{bd3}) is a system with hopping up to two lattice spacings, as well as a non-Hermitian term resulting from dissipation and/or driving.

{\it Fermi ribbon knots.---}
The nodal structure in Fig. \ref{threefoil} (b) exhibits a higher degree of complexity than those of Fig. \ref{nodes}, as the intersection of $\eta,\;\nu$ now must encode a trefoil knot. Applying a standard rational map ansatz \cite{Sutcliffe3001} 
we can generate these from the following construction: Assume a continuous function $f(\bk)$ that interpolates from $\pi$ to $0$ in the Brillouin zone according to
\bea
f:\;\bk\to \mathbb{R}, \; f(\bk=0)=\pi,\; f(\bk\in \partial b_z)=0, \label{f}
\eea
where $b_z$ is the Brillouin zone and $\partial b_z$ denotes its boundary. 
Then we may use this profile function to construct a map  $\mathbb{R}^3 \to \mathbb{C}^2$ according to
\bea
(Z_1,Z_0)=\Big([k_x+ik_y]\frac{\sin f}{|\bk|},\cos f+i\frac{\sin f}{|\bk|}k_z  \Big).\label{Zi}
\eea
Here we note that if $f$ is linear in $|\bk|$ for small momenta, then $\sin [f]/|\bk|\sim \text{constant}$ close to the origin so that Eq. (\ref{Zi}) defines a continuous function.  
Then the $(a,b)$ torus knot is given by the equation 
\bea
q(\bk,a,b)=Z_1^a+Z_0^b=0, \label{knotEq}
\eea
where $(a,b)=(2,3)$ corresponds to the trefoil knot.
To generate a knot-shaped intersection of $\eta,\nu$ we must thus encode Eq. (\ref{knotEq}) in the $\bd$ vector. For example, we may take
\bea
\bd= e_x q(\bk,a,b) +e_y p(\bk)+e_z \big[q'(\bk,a,b)-c+i\epsilon\big],\;\;\label{dKnot}
\eea
where $q,\;q'$ are knots of the same topology but not generally of exactly the same shape, and $p(\bk)$ denotes a perturbation to the $y$ component, which is assumed to be small. Taking $c=\epsilon=0.05$, $q'(\bk,a,b)=q(1.1\bk,a,b)$, and $(a,b)=(2,3)$ we obtain the nodal structure displayed in Fig. \ref{threefoil} (b). The explicit form of the profile function [Eq. (\ref{f})] used here is $f(\bk)=\theta(\pi-|\bk|)(\pi -|\bk|)$, where $\theta$ is the Heaviside function.

{\it Discussion.---}
In this work we have shown that the band structure of non-Hermitian systems 
admits generic nodal lines of nontrivial topology that are necessarily associated with open Fermi surfaces
which we dub Fermi ribbons. These provide novel and diverse higher-dimensional generalizations of Fermi arcs that are unique to three-dimensional non-Hermitian systems.    
This in turn has far-reaching implications for the topological characterization of the band structure. Once the nodes in the spectrum become linelike, we are forced to invoke knot theory to classify the resulting band touching points.  
Furthermore, in Hermitian systems in three dimensions, the generic band touching points are Weyl nodes that are associated with monopoles of Berry flux. The addition of non-Hermitian terms generally splits these into rings, yet not all exceptional lines originate in Weyl nodes. It is in principle possible to create a gapless system by adding a non-Hermitian term to an insulator. Thus, we must distinguish between exceptional lines that have topological origin and their trivial counterparts which can be shrunk to zero size and gapped out. 

Despite a superficial resemblance, nodal loops and links in Hermitian systems \cite{2016Natur.538...75B,PhysRevB.96.041102,PhysRevB.96.041103,PhysRevB.96.081114,PhysRevB.96.201305,PhysRevLett.119.147001} are not generic and therefore susceptible to arbitrarily small symmetry-breaking perturbations. Also, the preceding discussion of the relationship between defectivities and open Fermi ribbons reveals a dramatically different electronic structure.

In addition, let us note that if particle-hole symmetry is broken in Eq. (\ref{H}), then it is still possible to define the surfaces Eq. (\ref{Si}), though their energies are now shifted correspondingly, so that $S_1$ does not generally correspond to the Fermi surface. 
Yet, even in this case, the construction Eq. (\ref{Si}) implies that the nodal structure can be understood in terms of an open orientable surface $S_1$, whose boundary corresponds to defectivities in the theory, and so the implications for the topology of the exceptional lines do not rely on particle-hole symmetry.

Intriguingly, we find that it is possible to realize nodal lines of a highly nontrivial topology from relatively simple tight-binding models that are within reach of existing experimental techniques. A particularly promising platform for the observation of these phenomena is photonic systems \cite{topFot} in which both Weyl \cite{Lu622} and line-node \cite{10.1038/nmat4811} semimetal band structures have been emulated recently. 
In this context we note that adding a simple non-Hermitian term like Eq. (\ref{dissipation}), which has already been implemented in one-dimensional photonics systems, to a photonic Weyl node described by $H_{\rm eff}=v\mathbf k\cdot \sigma$ leads to an exceptional ring with radius $\epsilon/v$ and an accompanying open Fermi surface. An example of how bulk Fermi arcs can be realized in coupled resonator arrays is given in \cite{2018arXiv180508161M}.
Other promising platforms for realizing these phenomena include acoustic systems as well as 
 optically trapped ultra-cold atomic gases \cite{PhysRevLett.118.045701}. As alluded to in the Introduction, effective non-Hermitian models can also arise in electronic systems, either as a consequence of disorder in Weyl semimetals \cite{PhysRevB.97.041203}, or due to the finite lifetime of quasiparticles in heavy fermion systems \cite{2018arXiv180501172Y}, although these realizations would naturally offer less control over the dispersion. 
 
Although the race to first realize exceptional links and Fermi ribbons in experiment is still open, the recent dramatic progress in the field of Weyl semimetals, which is now turning into an integral part of our understanding of a range of seemingly very different materials, suggests that the robustness of generic band crossings will provide a main guiding principle in our understanding also of materials and metamaterials governed by non-Hermitian effective Hamiltonians.

Note added: During the editorial process the first observation of an exceptional ring in a 3D photonic crystal was reported in \cite{2018arXiv180809541C}.

{\it Acknowledgments.---}
We acknowledge discussions with Egor Babaev, Flore Kunst, and Marcus St\aa lhammar. This work was supported by the Swedish Research Council (VR) and the Wallenberg Academy Fellows program of the Knut and Alice Wallenberg Foundation.

\bibliography{biblio}

\end{document}